\documentclass[11pt]{article}

    \makeatletter
    \renewcommand\section{\@startsection {section}{1}{\z@}%
                                       {-3.5ex \@plus -1ex \@minus -.2ex}%
                                       {2.3ex \@plus.2ex}%
                                       {\normalfont\fontfamily{phv}\fontsize{16}{19}\bfseries}}
    \renewcommand\subsection{\@startsection{subsection}{2}{\z@}%
                                         {-3.25ex\@plus -1ex \@minus -.2ex}%
                                         {1.5ex \@plus .2ex}%
                                         {\normalfont\fontfamily{phv}\fontsize{14}{17}\bfseries}}
    \renewcommand\subsubsection{\@startsection{subsubsection}{3}{\z@}%
                                        {-3.25ex\@plus -1ex \@minus -.2ex}%
                                         {1.5ex \@plus .2ex}%
                                         {\normalfont\normalsize\fontfamily{phv}\fontsize{14}{17}\selectfont}}
    \makeatother
	
	\usepackage{amsmath}
	\usepackage{graphicx}
	\usepackage{enumerate}
	\usepackage{url} 

\usepackage[utf8]{inputenc}
\usepackage[T1]{fontenc}
\usepackage{hyperref}
\usepackage{xcolor}
\usepackage{amssymb}
\usepackage{url}            
\usepackage{booktabs}       
\usepackage{amsfonts}       
\usepackage{nicefrac}       
\usepackage{microtype}      
\usepackage{lipsum}
\usepackage{lmodern}
	
	\usepackage{amsfonts, amsthm, latexsym, amssymb}
	\usepackage{lineno}
	
\begin{document}

  \title{\bf Theoretical mapping of interaction between alkali metal atoms \\
  adsorbed on graphene-like BC$_3$ monolayer}
  \author{Kazem Zhour $^a$ and Andrei Postnikov $^a$ \\
			$^a$ LCP-A2MC Universit\'e de Lorraine 1 Bd Arago, F-57078 Metz Cedex 3, France \\
      \small Corresponding authors emails: \tt{kazem.zhour@gmail.com}; \tt{andrei.postnikov@univ-lorraine.fr} \\ }
  \date{}
  \maketitle

	\begin{abstract}
    First-principles calculations using density functional theory and 
    two methods in comparison, Quantum ESPRESSO and {\sc Siesta}, 
    are done on large supercells which describe different placements of two identical adsorbed 
    alkali metal atoms (of either Na, or K species) on the monolayer of boron carbide BC$_3$. 
    The energy of single-atom adsorption 
    over the center of C$_6$ ring, of the C$_4$B$_2$ hexagon and over a boron atom
    have been preliminarily estimated, the effect of applying the Grimme D2 correction 
    on the adsorption characteristics evaluated, and the comparison of
    these results with available data discussed.
    The interaction of two identical Na or K atoms
    adsorbed at close enough distances (less than ${\simeq}10$~{\AA}) 
    is negligible if the adsorption occurs at the opposite sides of the BC$_3$ layer,
    but creates a steep repulsive potential at distances less than ${\simeq}8$~{\AA}
    if both atoms are adsorbed on the same side of the monolayer.
    Relaxation patterns resulting from the two K atoms being trapped at adjacent adsorption sites 
    in the lattice are explained. 
    The results suggest that the density of adsorbed K atoms on BC$_3$ can be 
    interestingly high.
	\end{abstract}
			
	\noindent%


\section{Introduction}
\label{sec:intro}
Boron carbide (BC$_3$) recently emerged \cite{SSC136-22} as an interesting two-dimensional (2D) material, 
in its structure and some properties close to graphene, with the difference that
its single layer is semiconducting and not semi-metallic. 
Certain expectations are related with using this material functionalized with metals,
for so different purposes as alkali metal storage in batteries \cite{Carbon129-38},
enhanced sensibilisation for gas sensoring \cite{ApplSurfSci427-326},
and hydrogen storage \cite{JHydrogEne44-7354}.
These lines of research brought about recent works of first-principles simulation,
aimed at functionalisation of this material with adsorbed alkali metal atoms.
Zhao \emph{et al.} \cite{ChinChemLett32-900}
probed different placements of Li, Na and K atoms over different sites of BC$_3$,
compared the corresponding adsorption energies, and traced the energy profiles
across the barriers between adjacent adsorption sites. They used the VASP package on $3{\times}3$ supercells 
(of primitive cells with 2 formula units)
and generalized gradient approximation for the exchange-correlation after the prescription
of Perdew--Burke--Erznerhof (GGA-PBE) \cite{PRL77-3865,PRL78-1396}.
Naqvi \emph{et al.} \cite{ApplSurfSci512-145637} used the same calculation method
(on a smaller, $2{\times}2$ supercell) to probe the docking of the same alkali metals
and moreover of the alkaline earth metals, with subsequent adsorption of 
CH$_4$, CO$_2$ and CO molecules at the metal sites.
Bafekry \emph{et al.} \cite{JAP126-144304} explored different placements and estimated corresponding 
adsorption energies for a long list of adatoms and small molecules, in a series of calculations
using OpenMX and {\sc Siesta} methods with GGA-PBE, for $2{\times}2$ supercell.
All these studies, anticipating the importance of accounting for dispersion
interactions in the study of the systems in question, included semiempirical D2 correction 
after Grimme \cite{JCompChem27-1787}.

Whereas these studies permit to provide some consistent 
idea of how would the alkali metal atoms place themselves onto, and diffuse over,
the BC$_3$ layer, we find it interesting to address a hitherto unexplored issue
of interaction of adsorbed atoms at the surface. This may help to parametrise
this interaction to adequately model the diffusion, and/or to estimate
maximal density of atoms to be placed on the surface, which may be helpful 
for the above discussed issues, i.e., battery electrodes or gas sensors.
In view to probe this interaction, we considered supercells of larger size that
in the mentioned previous studies, and decorated with two (identical) alkali metal atoms.
Specifically, we studied K--K and Na--Na interactions.
Two calculation methods have been used in comparison: {\sc Siesta} for principal
large-size calculations and Quantum ESPRESSO for smaller-size benchmarks
(including the Li adsorption as well, to this end).
The calculation details are given in Section~\ref{sec:calcul}.
We start with a short discussion of pristine BC$_3$ and the single-atom adsorption on it
in Section~\ref{sec:BC3+Me}, checking ourselves against the previous results, 
and come to novel results concerning interaction in Section~\ref{sec:BC3+2K}.
The conclusions are summarized in Sec.~\ref{sec:conclu}.

\section{Calculation methods and technical details}
\label{sec:calcul}
We used two first-principles DFT methods in comparison, the plane-wave pseudopotential
Quan\-tum ES\-PRES\-SO (QE) \cite{JPCM21-395502,JPCM29-465901}
and numerical orbitals pseudopotential {\sc Siesta} \cite{JPCM14-2745,JChemPhys152-204108}.
Exchange-correlation was treated within the GGA, using 
the PBE parametrisation \cite{PRL77-3865,PRL78-1396}. 
{\sc Siesta} is generally expected to be more efficient than planewave methods
in treating large open low-coordinated systems, but, at the same time, 
potentially sensitive to the ``quality'' of fixed basis functions adopted. 
After some tests, we found it sufficient to use double-zeta basis functions for all atoms,
including notably $3p$ as a valence state for potassium. 
For sodium, the choice of the valence configuration
for the construction of pseudopotential, namely, the attribution of Na$2p$ states 
(situated in a free atom at ${\sim}\,-30$~eV) either to valence states or to the core
was not so obvious, since an accurate treatment of such semicore states
may be essential for correct grasping of fine total-energy trends.
Both cases have been tested.
The pseudopotentials used were norm-conserving, generated according to the
Trouiller -- Martins scheme \cite{PRB43-1993} for the following free-atom configurations
(the cutoff radii in Bohr being indicated in parentheses for each $l$-channel) : \newline
K $4s^1(3.14)\;3p^6(1.83)\;3d^0(3.14)\;4f^0(2.54)$, 
Na $3s^1(2.30)\;2p^6(2.30)\;3d^0(2.30)\;4f^0(2.30)$ and \newline
Na $3s^1(2.83)\;3p^0(2.83)\;3d^0(3.13)\;4f^0(3.13)$, 
B $2s^2(1.74)\;2p^1(1.74)\;3d^0(1.74)\;4f^0(1.74)$, \newline
C $2s^2(1.54)\;2p^2(1.54)\;3d^0(1.54)\;4f^0(1.54)$. 
The {\tt PAO.EnergyShift}
parameter in {\sc Siesta} was set to 0.015 Ry, that resulted in the maximal
extension of basis functions of 5~{\AA}. 

In QE calculations, the kinetic energy and the 
charge density cutoffs were set to 50 Ry and 400 Ry, respectively.
The convergence threshold for self-consistency in energy was set to $10^{-7}$~Ry.
Ultrasoft pseudopotentials have been used with QE.

A $8{\times}8{\times}1$ undisplaced $\mathbf{k}$ mesh was employed
in electronic structure calculations for pristine BC$_3$, with appropriate reduction in larger supercells.
The supercell ``thickness'' in the slab geometry (i.e., the se\-pa\-ra\-tion between 
periodically repeated BC$_3$ monolayers) was set at 30~{\AA}, sufficient to exclude spurious interaction 
of ad\-atoms with substrate across the vacuum layer, and to provide a reference ``no interaction''
energy of monolayer plus isolated atom when the latter was removed to the maximal distance (15~{\AA}).

\section{Pristine BC$_3$ and adsorption of alkali metals on it}
\label{sec:BC3+Me}
Planar Boron carbide, described in Ref.~\cite{SSC136-22}, is characterized by a honeycomb planar lattice
like that of graphene, in which one carbon atom out of four (equivalently,
two atoms within a $2{\times}2$ graphene supercell) are substituted by boron. 
In principle, one can imagine different
relative placements of B atoms on the honeycomb lattice, corresponding to 
para-, ortho- and meta-isomers (referred to as types A, B and C in Ref.~\cite{ModElecMat3-91}).
However, the para-isomer is singled out by its more regular and symmetric arrangement in which
every carbon atom has exactly one boron neighbour, and there are no B--B bonds.
This isomer retains the hexagonal (super)structure,
which makes a pattern of C$_6$ (perfect) hexagons with the C$-$C bond lengths $d_{\rm C-C}{\approx}1.4$~{\AA}
and C$_4$B$_2$ (distorted) hexagons around the perfect ones, with the C$-$B bond lengths
$d_{\rm C-B}{\approx}1.6$~{\AA}. 
The lattice parameter of this hexagonal unit cell which hosts two formula units is 
$a=\sqrt{3}(d_{\rm C-C}+d_{\rm C-B})$. 
Differently from perfect infinite graphene which has a zero gap, BC$_3$ has
the band gap of 0.65~eV (in a single sheet). A summary of previous data
concerning pristine BC$_3$, along with our present results, is given in Table~\ref{tab:BC3-pure}.   

\begin{table}[b!]
\caption{\label{tab:BC3-pure}Calculated properties of single-layer BC$_3$.}
\begin{center}
\begin{tabular}[htbp]{@{}llr@{.}lr@{.}lr@{.}lr@{.}l}
\hline
 Method & Ref. & \multicolumn{2}{c}{$d_{\rm C-C}$ ({\AA})} & 
                 \multicolumn{2}{c}{$d_{\rm C-B}$ ({\AA})} & \multicolumn{2}{c}{$a$ ({\AA})} & 
                 \multicolumn{2}{c}{$E_{\rm gap}$ (eV)} \rule[-4pt]{0pt}{15pt}  \\
\hline
{\sc Siesta} GGA & present work               & 1&428 & 1&564 & 5&183 & 0&67 \\
QE GGA           & present work               & 1&421 & 1&564 & 5&170 & 0&51 \\
VASP GGA         & \cite{Carbon149-733}       & 1&422 & 1&565 & 5&174 & 0&62 \\
VASP GGA         & \cite{ChinChemLett32-900}  & 1&42  & 1&57  & 5&17  & 0&66 \\
{\sc Siesta} GGA & \cite{JAP126-144304}       & 1&422 & 1&562 & 5&17  & 0&7  \\ 
\hline
\end{tabular}
\end{center}
\end{table}

The works aimed at the study of alkali metal adsorption on BC$_3$ from first principles 
are not numerous. Understandably, the possible adsorption sites probed in
previous works (notably Zhao \emph{et al.}\cite{ChinChemLett32-900})
are over the center of one or another hexagon, over the C$-$C or
C$-$B bonds, or over the one or the other atom species. Zhao \emph{et al.} argued
(with respect to Li, Na and K on BC$_3$, which we could confirm, see below),
that only three of these sites, over the both hexagons and atop a boron atom,
emerge as local energy minima for adsorption. 

There seems to be a consensus
among the calculations done that the hollow site over the center of the C$_6$ hexagon,
referred to as HC in the following, is the preferential one, closely followed by the hollow site
over the center of the C$_4$B$_2$ hexagon, referred to as HB. 
With respect to the magnitudes of the adsorption energy, however, there seems to be
a controversy we show in Table \ref{tab:BC3+K}.
We performed our calculations using sufficiently large supercell
(in terms of B$_2$C$_6$ unit cells) per adsorption atom and inspected several
possible sources of controversy. Without much details given in the papers cited,
good ``suspects'' could be the basis set superposition error (BSSE, Ref.~\cite{MolPhys19-553}),
potentially dangerous in methods with atom-centered basis functions 
(e.g., Ref.~\cite{JAP126-144304}), and the neglection of the magnetic state 
for an isolated potassium atom, both factors tending to overestimate the adsorption energy.
Our result shown in Table \ref{tab:BC3+K} is obtained from comparing the energies
of two sufficiently ``thick'' supercells of identical size, possessing the same 73 atoms: 
the relaxed system upon K adsorption, and the pristine BC$_3$ layer with the K atom
removed away from the layer at a distance that would preclude
the overlap of their basis functions. This does not fully removes the BSSE
(that could have been done by introducing ghost atoms), but substantially cancels
systematic errors. 

\subsection{{\sc Siesta} calculations for single adsorbed atoms (Na and K)}
\begin{figure}[!t]
\centerline{\includegraphics[width=0.9\textwidth]{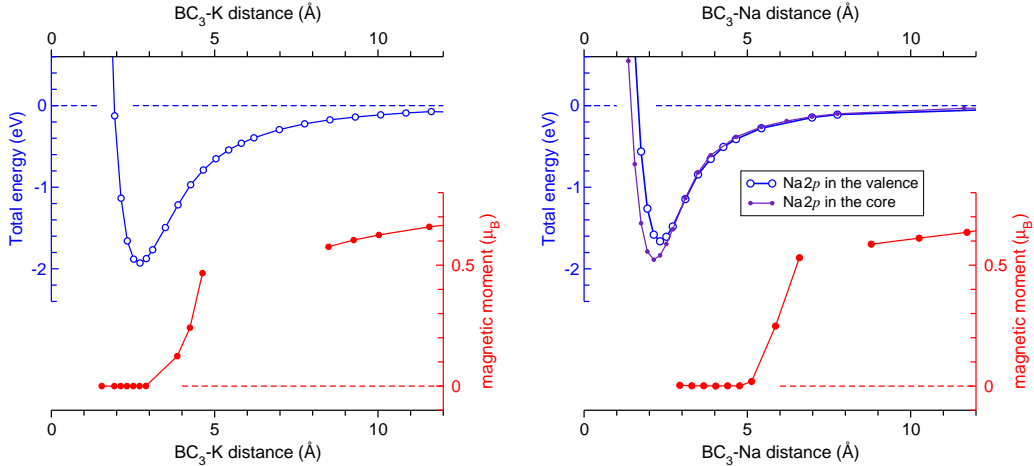}}
\caption{\label{fig:BC3+Me_mag}
Total energy (upper plots, blue curves, axes on the left) and magnetic moment (lower plots, 
red dots, axes on the right) 
as functions of the height of the alkali metal atom (left panel: K, right panel: Na) 
over the BC$_3$ layer. Calculations are done with lattice constant fixed at $a=5.17$~{\AA}
without structure relaxation. For Na, the results are shown for two norm-conserving 
pseudo\-poten\-ti\-als, for the cases of Na$2p$ states treated as valence states or attributed
to the core. See text for details.
}
\end{figure}

A smooth variation of the total energy and magnetic moment on approaching 
an alkali metal atom (K and Na have been tested)
to the BC$_3$ sheet (without performing any structure relaxation and only converging
the electronic structure) is depicted in Figure~\ref{fig:BC3+Me_mag}.
For the record, the lattice parameter in these trial calculations done with {\sc Siesta}
was fixed at $a=5.170$~{\AA}. 
The idea of this exercice was to inspect the sharpness of the 
potential energy minimum, and, in the case of Na, to test the stability of
results against the choice of pseudopotential, in the construction of which
the arguments could have been found to explicitly include the semicore Na$2p$ states
among the ``valence'' ones, or attribute them to the core. As can be concluded
from Figure~\ref{fig:BC3+Me_mag} (right panel), the difference in the potentiel energy profiles
stemming from these two choices is acceptably small yet noticeable. For the subsequent study
of Na--Na interactions, we ``pragmatically'' opted for using the ``small core''
Na pseudopotential with $2p$ treated as valence state. This choice results
in more accurate estimation of equilibrium atom-to-monolayer distance and yields a better
pronounced discrimination between the adsorption energies of Na and K atoms,
which seems to be consistent with general trend found by other calculations
listed in Table~\ref{tab:BC3+K} (whatever would be the sources of error
in reported \emph{absolute} energy values).
The depth of the total energy profile gives a fair estimate of the adsorption energy.
The magnetic moment exhibits instability throughout the range of distances 3 to 7~{\AA}
for K, or 5 to 8~{\AA} for Na, as the outer $s$ state of the corresponding alkali metal,
which carries an unpaired spin in a free atom, starts to interact with the wavefunctions
of the BC$_3$ layer. The corresponding part is therefore not shown in the magnetic moment plots
in Figure~\ref{fig:BC3+Me_mag}. Upon adsorption
and after atomic and electronic relaxation, the magnetic moment is fully dissolved
(differently e.g. from the situation with graphene, on which the adsorbed K atom
retains its magnetic moment).\footnote{%
The asymptotic value of the magnetic moment corresponding to 
large distances in Figure~\ref{fig:BC3+Me_mag}
does not reach $1~{\mu}_{\rm B}$ -- probably due to a purely technical drawback
that very fine $\mathbf{k}$ and energy mesh is needed to cope with very narrow $s$-states
energy levels.}

\begin{table}[!b]
\caption{\label{tab:BC3+K}
Calculated parameters of alkali metal atoms adsorbed at C$_6$ hollow site of BC$_3$.
$E_a$: adsorption energy (defined positive, i.e., corresponding to energy gain on adsorption, in all cases);
$h$: the height of the adsorbed atom over the BC$_3$ plane. See text for detail.}
\begin{center}
\begin{tabular}[htbp]{lll c ccc c ccc}
\hline
        &&&& \multicolumn{3}{c}{$E_a$ (eV)} && \multicolumn{3}{c}{$h$ ({\AA}) \rule[-4pt]{0pt}{15pt} } \\
        \cline{5-7} \cline{9-11}
 Method & Ref. & Supercell && Li & Na & K & \rule[-4pt]{0pt}{15pt} & Li & Na & K \\
\hline
OpenMX or {\sc Siesta} (GGA)$^{\dagger}$ & \cite{JAP126-144304} & $2{\times}2{\times}1$ &&
0.81 & 1.34 & 0.82 && 2.31 & 2.23 & 2.57 \\ 
VASP (GGA)$^{\dag}$        & \cite{ApplSurfSci512-145637}       & $2{\times}2{\times}1$ &&
5.13 & 4.47 & 4.72 && 1.74 & 2.15 & 2.53 \\
VASP (GGA)$^{\dag}$        & \cite{ChinChemLett32-900}          & $3{\times}3{\times}1$ &&
1.26 & 1.16 & 1.75 && 1.77 & 2.18 & 2.58 \\
{\sc Siesta} (GGA) & present work$^{\ast}$                      & $3{\times}3{\times}1$ &&
     & 1.53 & 1.66 &&      & 2.18 & 2.68 \\
QE (GGA)$^{\ddag}$         & present work$^{\ast}$              & $3{\times}3{\times}1$ &&
2.56 & 1.85 & 2.31 && 1.73 & 2.26 & 2.63 \\
\hline
\end{tabular}
\\
$^{\dag}$with Grimme-D2;
$^{\ddag}$for results with Grimme-D2 correction, see next table. 
$^{\ast}a$=5.17~{\AA} fixed.
\end{center}
\end{table}

The results obtained in the course of relaxing atomic positions within a supercell 
of a fixed size, corresponding to the lattice parameter $a=5.17$~{\AA}
(as optimized in QE calculation), by comparing the total energies from ``adsorbed atom'' 
and ``distant atom'' situations within the same calculation setup, are shown in Table~\ref{tab:BC3+K}.
For comparison, using as reference values the total energies of the BC$_3$-only supercell and
the ``single-atom-only'' supercell instead of the ``BC$_3$ plus distant atom'' one
yields the adsorption energies of ${\simeq}2.1$~eV, for the case of potassiuim.

\subsection{Adsorption energies calculated with QE; role of dispersion interactions}

Independently, we calculated the adsorption energy of $M$=(Li, Na, K) atoms at HC, HB and AB
sites by the QE method, also for the $3{\times}3{\times}1$ supercell,
referring in this case to the total energy differences of the kind
(Supercell with adsorbed $M$) $-$ [(Supercell without $M$) + (Single $M$ atom in the box)],
which approach is less problematic when using a planewave-basis code not prone to BSSE.
In view of expected non-negligible effect of dispersion interactions
(not included in conventional PBE calculations) on the adsorption characteristics,
we simulated it (the effect) on the total energy within broadly used
semiempirical Grimme's G2 correction\cite{JCompChem27-1787}; the equilibrium geometries 
were affected accordingly. The results (adsorption energies and
equilibrium heights of adsorbed atoms over the BC$_3$ plane) are shown in Table~\ref{tab:QE_E+z}, 
the net effect of the G2 correction to the GGA results being placed in brackets.

The calculated adsorption energy values are consistent with 
those from the recent work by Zhao \emph{et al.}\cite{ChinChemLett32-900}
by the order of magnitude and in an observation that the adsorption at HC and HB sites
is very competitive (being in all cases stronger in HC), both corresponding energies 
being larger than that at the AB site. The exact values are systematically, albeit slightly,
larger in our QE calculation.
We note in passing that the calculated adsorption energies reported 
by Naqvi \emph{et al.}\cite{ApplSurfSci512-145637}
are largely overestimated by roughly a factor of two while those reported by Bafekry
\emph{et al.}\cite{JAP126-144304} are largely, and unevenly, underestimated.

We note however that the variation of adsorption energies over different 
adsorption sites, which is the smallest ($\simeq\,0.13$~eV) for K, larger ($\simeq\,0.2$~eV) 
for Na and the largest ($\simeq\,0.5$~eV) for Li, is in perfect agreement between our study and  
Ref.~\cite{ChinChemLett32-900} (see Figure~5 therein). This may be due to the fact that
systematic errors potentially present in calculations of adsorption energies by this or that
method are practically cancelled in the estimation of ``barrier heights'' between adsorption
sites. 

The equilibrium positions of adsorbed atoms over the monolayer are in excellent 
agreement with almost all previous calculations (see Table~\ref{tab:BC3+K}).
As is seen from Table~\ref{tab:QE_E+z}, equilibrium distances are in general only weakly affected
by inclusion of the Grimme's correction, hence pro\-bably more ``reliable'' and reproducible character 
of these calculated values.

The hence discussed relations between energies of different species
at different adsorption sites underline a special standing of potassium among alkali metal atoms 
in that its potential energy landscape over the BC$_3$ layer is particularly flat, which may imply 
a good mobility. This comes 
with an appreciable adsorption energy, so -- simplifying -- potassium easily sticks to BC$_3$
\emph{and} easily rolls over it.

In the next section, we come to the analysis of interaction between adsorbed alkali metal atoms.

\begin{table}[!t]
\caption{\label{tab:QE_E+z}
Adsorption energy (defined positive for energy gain) and height over BC$_3$ layer for Li, Na and K atoms
in three symmetric positions, according to
QE calculations with PBE (upper line in each block for a given atom) and with Grimme-D2 correction 
included (lower line; the value in parentheses gives the difference with respect to the PBE value).}
\begin{center}
\begin{tabular}{*{2}{c*{3}{c@{$\!\;$}c}}}
\hline
    & \multicolumn{6}{c}{Adsorption energy (eV)} & \rule[-2pt]{0pt}{12pt} 
    & \multicolumn{6}{c}{Height over layer ({\AA})} \\
       \cline{2-7} \cline{9-14} 
    & \multicolumn{2}{l}{HC} & \multicolumn{2}{l}{HB} & \multicolumn{2}{l}{AB} & \rule[-2pt]{0pt}{12pt} 
    & \multicolumn{2}{l}{HC} & \multicolumn{2}{l}{HB} & \multicolumn{2}{l}{AB} \\
\hline
    & 2.56 &           & 2.47 &           & 2.09 &           &&  
      1.73 &           & 1.72 &           & 1.91 &           \\*[-6pt]
 Li \\*[-6pt]        
    & 2.91 & ($+$0.36) & 2.85 & ($+$0.39) & 2.41 & ($+$0.32) &&
      1.80 & ($+$0.08) & 1.73 & ($+$0.06) & 1.73 & ($-$0.18) \\
\hline
    & 1.85 &           & 1.83 &           & 1.69 &           &&
      2.26 &           & 2.22 &           & 2.28 &           \\*[-6pt]
 Na \\*[-6pt]        
    & 2.23 & ($+$0.38) & 2.22 & ($+$0.40) & 2.02 & ($+$0.33) &&
      2.28 & ($+$0.02) & 2.23 & ($+$0.01) & 2.23 & ($-$0.05) \\
\hline
    & 2.31 &           & 2.25 &           & 2.20 &           &&
      2.63 &           & 2.62 &           & 2.65 &           \\*[-6pt]
 K \\*[-6pt]        
    & 2.61 & ($+$0.29) & 2.56 & ($+$0.30) & 2.48 & ($+$0.28) &&
      2.63 & ($+$0.00) & 2.62 & ($+$0.00) & 2.58 & ($-$0.07) \\
\hline
\end{tabular}
\end{center}
\end{table}

\section{Interaction of alkali metal atoms adsorbed on BC$_3$}
\label{sec:BC3+2K}

\subsection{Spatial map of $M$--$M$ interaction energies on the BC$_3$ surface}

\begin{figure}[!b]
\begin{center}
\includegraphics[width=0.8\textwidth]{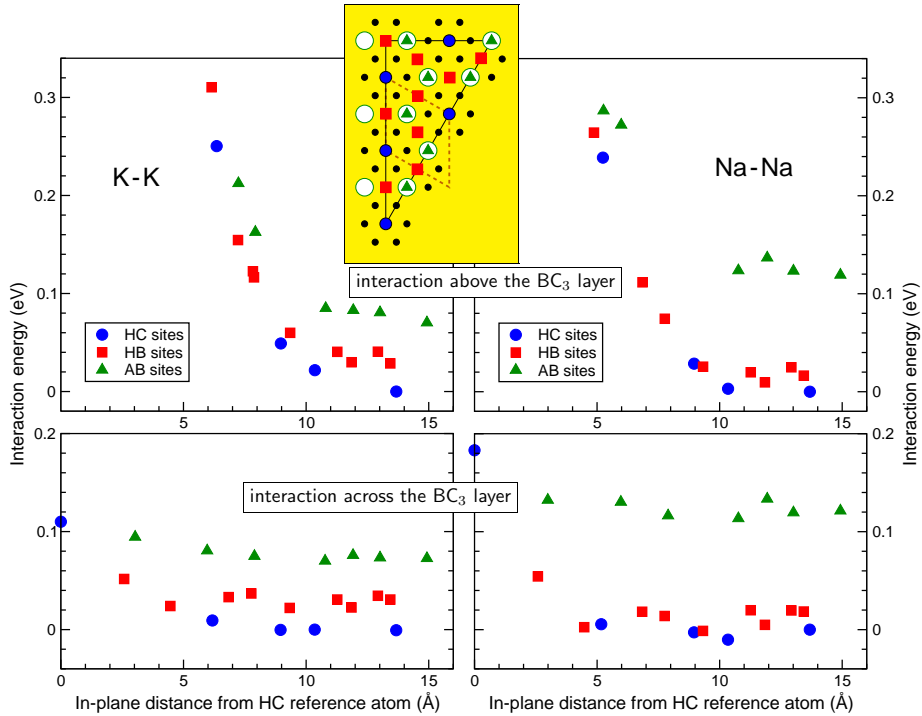}
\end{center}
\caption{\label{fig:Me-Me_map}
Map of interaction energies between alkali metal atoms 
(two panels on the left refer to K--K, two on the right to Na--Na),
adsorbed at different sites 
of the BC$_3$ lattice (marked by circles, squares and triangles in the ``irreducible part''
of the $5{\times}5$ supercell schematically shown in the inset), 
and the reference atom in the HC position (the bottom blue point in the inset).
HC positions (over C$_6$ hexagons), HB positions (over C$_4$B$_2$ hexagons)
and AB positions (over B atoms) are indicated by different symbols.
The two upper panels refer to the interactions between atoms adsorbed at the same side
of the BC$_3$ plane, the bottom panels -- between atoms adsorbed
at the opposite sides, in both cases -- as function of the in-plane
distances between the adsorption sites. Positions included in the mapping are marked
by symbols, as in the main plot. A primitive cell of BC$_3$ is 
outlined by a dashed contour in the inset.}
\end{figure}

In order to gain a substantial and reliable information concerning
the interaction between alkali metal atoms $M$ adsorbed on BC$_3$
(specifically, Na and K have been studied), we staged 
a series of simulations on large supercells containing two adsorbed atoms. 
One of them, the ``reference atom'', was placed in a (lowest-energy) HC
position over the center of a C$_6$ ring,
and the other, ``trial'' atom -- at different local-minima positions at different 
distances from the reference atom. Within the $5{\times}5$ supercell
(200 atoms), all non-equivalent adsorption positions have been explored
within the ``irreducible wedge'' (shown in inset of Figure~\ref{fig:Me-Me_map})
closest to a given HC site. This amounted to four HC positions, 
9 HB positions (over the center of C$_4$B$_2$ hexagons) and 7 AB positions
(atop of a boron atom), hence the positions confirmed (see previous Section)
to correspond to stable (local energy minima) adsorption sites.
Moreover, the con\-fi\-gu\-rations in which the second (trial) atom was adsorbed 
at the \emph{opposite} side of the BC$_3$ monolayer have been tested -- on all the HC, 
HB and AB sites, mentioned above,
plus directly at the reference site, in a mirror position from the reference atom.
In all cases, an unconstrained relaxation has been performed
(fixing however the supercell parameters), which yielded
the total energy and the equilibrium geometry. 

For $M$--$M$ distances starting from ${\sim}7${\AA} onwards, the relaxation patterns
are barely visible: the adsorbed atoms reside in their nominal positions,
so that only tiny modifications of interatomic distances come about.
The BC$_3$ layer in all cases looks extremely rigid and planar, without
a marked tendency towards bending or warping.
At shorter distances, the repulsion between the adsorbed atoms
pushes them out of symmetry positions; the examples of this will be discussed
below. In fact, the stabilisation of the trial atom in the nearest HB or AB position 
to the reference atom becomes impossible, because the atoms drift away 
from such configuration. This refers to the atoms adsorbed at the same side
of the CB$_3$ plane. The atoms adsorbed ``across the plane'' remain 
practically insensitive to the closeness of the reference atom,
exhibiting no perturbations in their relaxation pattern, and the energies
which are largely independent on the in-plane distance to the reference atom.

The ``interaction energies'' plotted in Figure~\ref{fig:Me-Me_map} are defined as
total energies from each calculations, expressed relative to the energy of supercell in which
the two atoms are at maximally separated HC positions, within the given type of placement 
(i.e., either both above, or one below the BC$_3$ layer), effectively setting the
``infinity limit'' within the given supercell size.
We note that these ``infinity limit'' energies
are not exactly equal for the trial atom being on the same side of the BC$_3$ plane
as the reference atom, or on the opposite side. In fact, the second energy
is lower by 0.03~eV for Na adsorption and by 0.05~eV for K adsorption.
This disparity (which would disappear be the atoms at
really sufficiently far distance) can be considered as a measure of
the effect of our limited supercell size.

Choosing such ``distant $M$--$M$'' limit of two adsorbed atoms
as the reference level leaves the \emph{adsorption energy} at a HC site out of
our grasp; however, this issue has been separately addressed in the 
previous section. However, the \emph{differences} between adsorption energies
at HC, HB and AB sites, or otherwise the barrier heights, are easily readable
from Figure~\ref{fig:Me-Me_map}, and are instructive to discuss.

We note first that for ``inverted'' adsorption of a trial atom
(cf. bottom panels in Figure~\ref{fig:Me-Me_map}), the energy
of its interaction with the reference one is practically independent
on interatomic distance, and follows the universal trend that HB positions
are, on the average, higher in energy by ${\simeq}0.03$~eV than the HC ones,
whereas the AB positions are higher than the HC by ${\simeq}0.08$~eV,
in case of K adsorption. As for the Na adsorption, the energies at HC and HB positions 
are almost identical within the ``numerical noise'', the HB ones being just
a bit higher, wereas the energies in AB positions are much higher, by ${\simeq}-0.12$~eV.
This simply passes the information about barrier heights (differences between
adsorption energies at different sites), which was already addressed 
in the previous section, see e.g. Table~\ref{tab:QE_E+z}.
However, here it appears not as a ``random'' number
but as a ``statistically credible'' result. We can again refer to the similarity
of the present estimation of barrier heights with those given by  
Zhao \emph{et al.} \cite{ChinChemLett32-900}, notably in Figure~5(c,d)
of this publication.

The bottom panel of Figure~\ref{fig:Me-Me_map} suggests 
that the alkali metal atoms ``do not see each other'' across the BC$_3$ plane
till the $M$--$M$ distance becomes really short, within the closest 
adsorption sites. Still, this does not bring about neither an ``asymmetric''
relaxation nor noticeable augmentation of the $z$-coordinate;
the adsorbed atoms remain well centered at their respective sites.
The ``double occupation'' of a HC adsorption site from both sides 
of the BC$_3$ plane, whereby the two K atoms remain stable at a distance 5.4~{\AA}
(hence at a regular height above the layer) 
is by far not so energetically unfavourable as it could have been
\emph{a priori} anticipated; it costs only 0.1~eV to bring a K atom
from a distant HC site onto this ``antipode'' reference site; for Na atoms
the corresponding energy is 0.18~eV.

The established hierarchy of barrier heights, revealed via
interaction energies being independent on the $M$--$M$ distance,
is also valid for adsorptions at the same side of the BC$_3$ plane 
(see upper panels of Figure~\ref{fig:Me-Me_map}), provided the $M$ atoms
are placed not closer than ${\simeq}10$~{\AA}. At shorter distances,
the energy barrier rapidly increases, forming a nearly rigid core 
of the radius of ${\simeq}5$~{\AA} for Na and ${\simeq}6$~{\AA} for K, irrespectively 
of the adsorption site concerned.

\subsection{Unusual relaxation patterns of closely placed adsorbed atoms}

\begin{figure}[!b]
\includegraphics[width=0.95\textwidth]{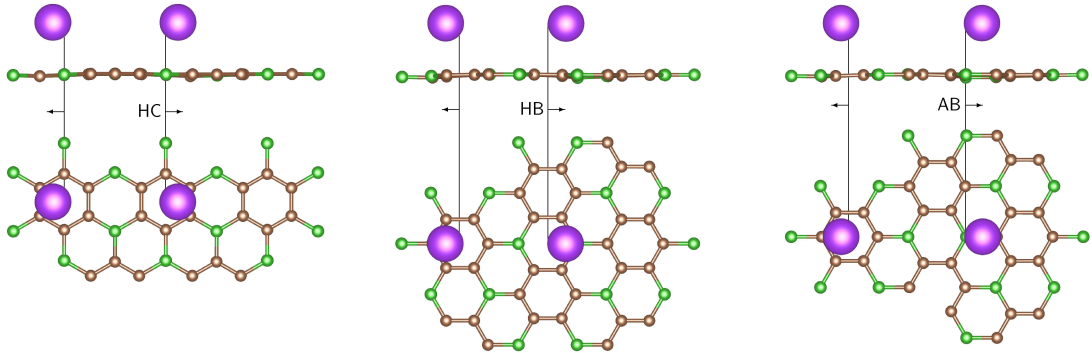}
\caption{\label{fig:relax}
Three relaxed structures (top row: side views, bottom row: top views) of trial K atoms initially placed
at HC, HB and AB positions close to the reference K atom 
(that in the left of each plot). The arrows indicate the movement
of the both K atoms apart.}
\end{figure}

The placement of trial atoms at adsorption sites close (within ${\simeq}8$~{\AA})
of the reference atom resulted in a noticeable repulsion of both atoms
in their relaxed configurations, as shown in Figure~\ref{fig:relax}
for the case of potassium.
The reference atom displaces in its ``nest'' over the C$_6$ ring
without however leaving the latter, whereas the trial atom could have been
pushed from its initial ``on top of B'' position onto that above the C--B
bond; see the case AB in Figure~\ref{fig:relax}. It was not possible
to stabilise the trial atoms in AB or HB positions closest to the reference
atom, since the interacting atoms were pushed apart into ``second-neigh\-bours''
configurations, shown in Figure~\ref{fig:relax}. This strengthens an idea of
``hard core'' repulsion potential which effectively prevents the potassium atoms
to come closer than about 6~{\AA}. This minimal distance corresponds to
both atoms residing over the second-neighbour C$_6$--C$_6$
or C$_6$--C$_4$B$_2$ hexagons, slightly pushed apart from
the respective centers, as shown in Figure~\ref{fig:relax}.
We note that the $z$-coordinates of potassium atoms are not much affected
by their relaxation ``situation''; also the warping of the BC$_3$ layer,
although noticeable around such ``strong'' relaxations, remains small
and difficult to make sense of.

\section{Conclusion}
\label{sec:conclu}
Summarizing, we found that alkali metal atoms, sodium and potassium,
adsorbed over the BC$_3$ monolayer
interact strongly repulsively at distances smaller than ${\simeq}10$~{\AA},
and otherwise remain relatively insensitive to the mutual presence, feeling
the ``usual'' landscape of potential barriers of ${\lesssim}0.1$~eV (case of K)
or ${\lesssim}0.15$~eV (case of Na)
in their movement over the BC$_3$ plane. A combination of relatively high
adsorption energy with respectively low barriers, especially prominent for potassium,
known already from early works, suggested already an interesting interplay
of sticking and mobility. Now we conclude that even relatively high
concentration of adsorbed K atoms might not prevent them from acting as
a highly mobile ``gas'' on the BC$_3$ surface. This may open perspectives
for further more realistic simulations and interesting applications.   

\medskip
\textbf{Supporting Information} \par 
which includes tables with (relaxed) interatomic distances and energies for all the adsorption 
sites (the data depicted in Figure~\ref{fig:Me-Me_map}) is available from the Wiley Online Library 
or from the authors.

\medskip


\end{document}